\documentclass{aastex}
\usepackage{emulateapj5}
\usepackage{epsfig}

\shorttitle{Magnetic Field Morphology of Orion-IRc2}
\shortauthors{Plambeck, Wright, \& Rao}

\newcommand{\vlsr}{V$_{\rm LSR}$}
\newcommand{\kms}{\mbox{km s$^{-1}$}}
\newcommand{\cucm}{cm$^{-3}$}
\newcommand{\jone}{\mbox{J=1--0}}
\newcommand{\jtwo}{\mbox{J=2--1}}
  
\begin{document}
\slugcomment{accepted for publication in ApJ}
 
\title{Magnetic Field Morphology of Orion-IRc2 \\
from 86~GHz SiO Maser Polarization Images}

\author{R.L.~Plambeck and M.C.H.~Wright}
\affil{Radio Astronomy Lab, University of California, Berkeley, CA 94720}
\email{plambeck@astro.berkeley.edu, wright@astro.berkeley.edu}
\and
\author{R.~Rao}
\affil{University of Chicago, 5720 South Ellis Ave., Chicago, IL 60637}

\begin{abstract}

In an attempt to probe the magnetic field morphology near the massive
young star Orion-IRc2, we mapped the linear polarization of its \jtwo\
SiO masers, in both the v=0 and v=1 vibrational levels, with $0.5''$
resolution.  The intense v=1 masers are confined to a narrow zone 40~AU
from the star.  Their polarization position angles vary significantly on
time scales of years.  For the v=1 masers the stimulated emission rate
$R$ is likely to exceed the Zeeman splitting $g\Omega$ due to any
plausible magnetic field; in this case the maser polarization need not
correlate with the field direction.  The much weaker v=0 masers in the
ground vibrational level lie \mbox{100--700~AU} from IRc2, in what
appears to be a flared disk.  Their fractional polarizations are as high
as 50\%.  The polarization position angles vary little across the line
profile or the emission region, and appear to be stable in time.  The
position angle, P.A.  = 80\degr, we measure for the \jtwo\ masers
differs by 70\degr\ from that measured for the \jone\ SiO transition,
possibly because of Faraday rotation in the foreground, Orion~A,
\ion{H}{2}\ region.  A rotation measure $RM = 3.3 \times 10^4$
rad~m$^{-2}$ is required to bring the \jtwo\ and \jone\ position angles
into concordance.  The intrinsic polarization position angle for both
transitions is then 57\degr, parallel to the plane of the putative disk. 
Probably the magnetic field threads the disk poloidally.  There is
little evidence for a pinched or twisted field near the star. 

\end{abstract}


\section{Introduction}

As an interstellar cloud collapses to form a star, the magnetic field
lines threading the cloud are expected to be dragged along with the
infalling gas, producing an hourglass-shaped field pattern centered on
the star (e.g., Galli \& Shu 1993).  In many theoretical models the
field is then wound up by rotation in a circumstellar disk, producing a
toroidal field component which collimates the outflow from the star into
two oppositely directed lobes; the outflow carries away angular
momentum, allowing further accretion onto the star \citep{Uchida85,
Pudritz86,Newman92}. 

So far the observational evidence for pinched or twisted fields near
young stars is scant, although there are possible detections toward W3
\citep{Greaves94}, Mon~R2 \citep{Greaves95}, and a number of other cloud
cores \citep{Greaves98}.  These observations used the linearly polarized
emission from circumstellar dust to probe the magnetic field morphology. 
Spinning dust grains tend to align their long axes perpendicular to the
magnetic field, so thermal emission from the grains is polarized
perpendicular to the field. 

Toward the Orion molecular cloud, \citet{Schleuning98} found evidence
for a large scale hourglass distortion of the magnetic field from dust
polarization observations at 100 and 350~$\mu$m.  The pinched region has
a radius of $\sim$0.5 pc (200\arcsec) and is centered 0.1 pc southwest
of the Kleinmann-Low nebula, a site of high mass star formation.  On
smaller scales, within the Orion-KL complex, \citet{Chrysostomou94}
found evidence for a twisted field within roughly 1500 AU (3\arcsec) of
the star IRc2, based on imaging polarimetry of the 2~$\mu$m S(1) line of
H$_2$.  The line emission, assumed unpolarized, originates from
shock-excited H$_2$ in the bipolar outflow from IRc2.  Absorption by
aligned grains in front of the outflow produces the polarization. 
\citet{Aitken97} also found evidence for a twisted field near IRc2 from
1.5\arcsec\ resolution polarization maps at 12.5~$\mu$m and 17~$\mu$m. 
Spectropolarimetry of the 8--13 and 16--22~$\mu$m bands was used to
separate the effects of absorption and emission, which lead to
polarizations parallel and perpendicular to the magnetic field,
respectively.  \citet{Rao98} obtained 4\arcsec\ resolution
interferometric maps of Orion-KL at wavelengths of 1.3 and 3.5~mm, where
the dust emission is certainly optically thin.  These maps showed an
abrupt 90\degr\ change in the polarization angle southeast of IRc2, in
agreement with the infrared data.  But because the region of anomalous
polarization was suspiciously coincident with the redshifted lobe of the
bipolar outflow from IRc2, Rao et al.  suggested that dust grains in
this zone might be aligned by a wind from the star, rather than by the
usual Davis-Greenstein mechanism.  If the grains are mechanically
aligned then thermal emission can be polarized parallel to the magnetic
field, so there is no need to conclude that the field is kinked. 

Unfortunately, polarized dust emission from IRc2 may not be a reliable probe
of the magnetic field near the central star because of confusion by dust
emission from the surrounding molecular cloud.  The stellar position is
coincident with radio source `I' \citep{Menten95}.  The IRc2 infrared
peak is offset $\sim$1\arcsec\ northwest of source I, while the
millimeter-wavelength peak is offset $\sim$1\arcsec\ southeast of
source I, toward the `hot core' clump.  The radio spectrum of source~I
itself from 8~GHz to 86~GHz is consistent with free-free emission from
ionized gas in a circumstellar shell; there is no evidence for excess
flux at 86~GHz which would be the signature of circumstellar dust
\citep{Plambeck95}. 

By contrast, SiO masers do unambiguously originate in gas very close to
source~I.  Intense masers in the v=1 vibrational level are confined to a
zone approximately 40 AU from the star \citep{Plambeck90,Menten95,
Greenhill98,Doeleman99}, while weaker masers in the ground vibrational
state are found 100--700~AU from the star, in an elongated,
bowtie-shaped region \citep{Wright95,Chandler95}.  Both the v=1 and v=0
masers are linearly polarized \citep{Barvainis84,Tsuboi96}.  Since the
maser polarization direction is expected, under many circumstances, to
be parallel or perpendicular to the magnetic field direction
\citep{Goldreich73}, SiO masers may provide the best probe of the
magnetic field morphology within 700~AU of source~I.  Indeed, from
early measurements showing a symmetric rotation of the polarization
angle across the double-peaked v=1 maser spectrum, \citet{Barvainis84}
concluded that the magnetic field direction rotated with azimuth in a
disk around the star. 

Here we discuss polarization measurements of both the v=0 and v=1 SiO
masers toward Orion-IRc2 obtained with 0.5\arcsec\ angular resolution,
$\sim$250~AU at the distance of Orion.  We find that the polarization of
the v=1 maser is time variable, casting doubt upon its usefulness as a
probe of the magnetic field.  However, the v=0 maser polarization
appears to be a good probe of the field morphology.  It provides little
evidence for pinched or twisted fields near the star. 
\section{Observations}

Observations were made between 1996 June and 2001 March with the BIMA
array.\footnote{Operated by the Berkeley-Illinois-Maryland Association
with support from the National Science Foundation.} Orion-KL was
observed using the A, B, and C antenna configurations, providing antenna
separations from 6.3~m to 1.8~km.  The correlator was configured to
allow simultaneous observations of the \jtwo\ SiO transitions in both
the v=1 and v=0 vibrational states, at frequencies of 86.243 and 86.847
GHz.  The velocity resolution was 0.34~\kms.  Single sideband system
temperatures, scaled to outside the atmosphere, ranged from 170~K to
350~K.  The data were calibrated and analysed using the MIRIAD software
package \citep{Sault95}. 

The receivers are sensitive to a single linear polarization, but are
equipped with movable quarter wave plates to observe right or left
circular polarization.  As described by \citet{Rao98}, each receiver was
switched between LCP and RCP with a Walsh function pattern in order to
sample all possible crosscorrelations (RR, RL, LR, LL) on each baseline
as rapidly as possible.  The integration time for each observation was
11.5 seconds.  The data were self-calibrated on the strongest v=1 maser
feature, using the LL and RR crosscorrelations, then averaged in 5
minute blocks to produce quasi-simultaneous dual polarization
observations. 

The quarter wave plates are designed for a frequency of~89 GHz, where
they have a leakage to the unwanted polarization of around 1\%.  At
86~GHz the leakage is about 3\%.  We calibrate the leakage by observing
a point source over a wide range of hour angle.  For altitude-azimuth
antennas the phase of a linearly polarized source rotates with the
parallactic angle, whereas the leakage terms are constant.  We used
observations of 3c273, 3c279, and the SiO v=1 maser itself to calibrate
the leakage.  The measured polarizations of both the v=1 and v=0 masers
are quite insensitive to the leakage correction: for the weakly
polarized v=1 line, the source is unresolved and the leakage averages
out over the wide range of parallactic angles observed, while for the
v=0 line the source polarization is much larger than the leakage. 

Images were made for a range of weightings of the uv-data.  With uniform
weighting the synthesised beam is $ 0.8'' \times 0.3 ''$ FWHM.  For the
final images we used robust weighting, deconvolved using the CLEAN
algorithm, and convolved with a circular $0.5''$ FWHM Gaussian beam.

\section{Results}

\subsection{The v=1 Masers}

\begin{figure}
\begin{center}
\vskip0.1in
\epsfig{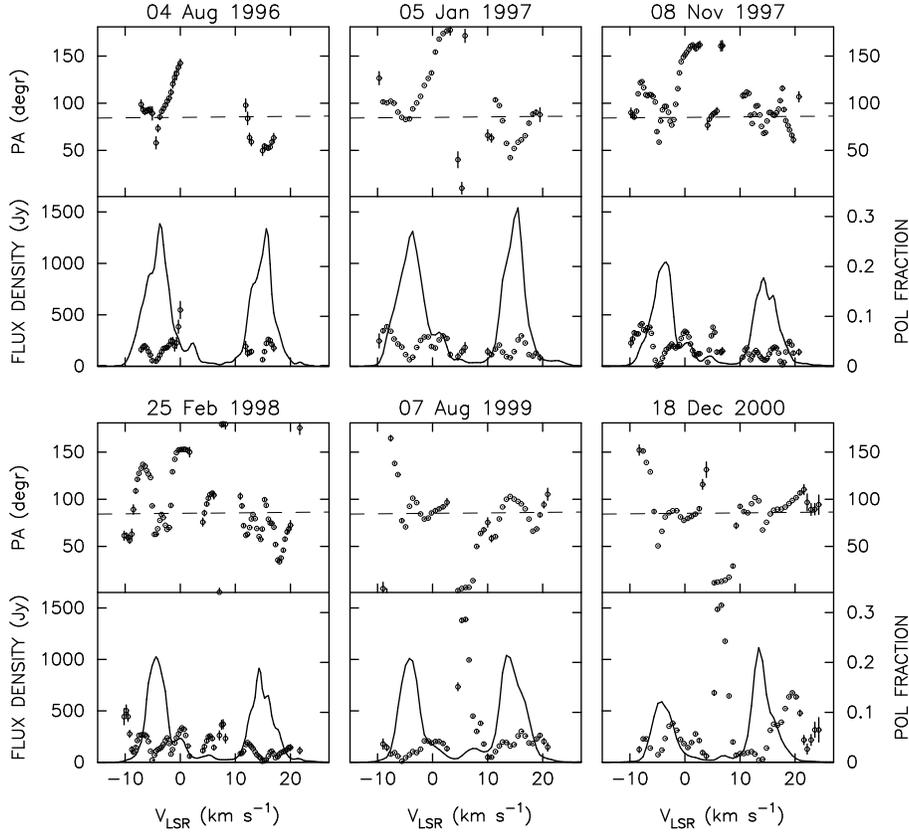}
\end{center}

\caption{Total intensity, fractional polarization, and position angle
for the v=1 \jtwo\ SiO masers (86.243 GHz) in Orion-IRc2, at 6 epochs. 
Fractional polarizations and position angles are plotted as points, with
$\pm 1\sigma$ error bars indicated.  A dashed line is drawn at P.A.  =
80\degr; this is the position angle of the v=0 masers.  }
\label{fig:f1}
\end{figure}

Figure~\ref{fig:f1} summarizes the observational results for the v=1
masers over a 3.5 year period.  For each of 6 epochs, we plot the total
intensity, fractional polarization, and polarization position angle
across the double-peaked spectrum.  All 3 quantities vary on time scales
of years.  The fractional polarization usually is less than 10\%, and
tends to be anticorrelated with maser intensity -- that is, the
strongest maser features typically have the lowest fractional
polarization.  The weak maser feature near \vlsr\ 7 \kms\ which appeared
in 1999 has an anomalously high fractional polarization of 30\%.  The
polarization position angles vary considerably across the line profiles,
although a crude median value is P.A.  $\sim$ 80\degr, shown by the
dashed line in Figure~\ref{fig:f1}.  Often the position angle ``kinks''
at the peak of a maser feature.  The symmetric pattern of polarization
position angles seen in the 1996 August data is similar in character to
that measured by \citet{Barvainis85} in 1981 June. 

\begin{figure}
\vskip0.1in
\begin{center}
\epsfig{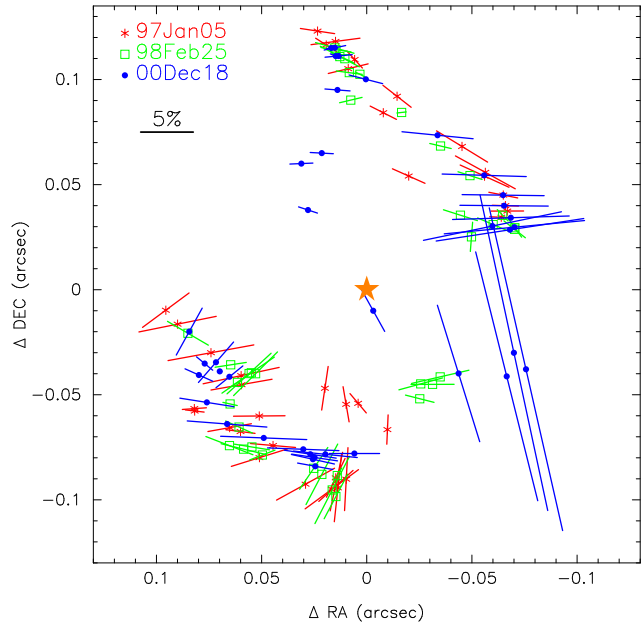}
\end{center}
\caption{Polarization vectors plotted at the fitted centroid positions
for the v=1 \jtwo\ SiO masers in Orion-IRc2, for 1 \kms\ velocity
channels.  Data for 3 epochs are overplotted to search for correlations
between centroid positions and polarizations.  Radio source~I, denoted
by a star, is located near the middle of the maser distribution.  The
arc of maser spots SE of source I corresponds to the blueshifted peak of
the maser spectrum (Figure~\ref{fig:f1}), while the arc of spots to the
NW corresponds to the redshifted peak.  Position uncertainties are of
order $\pm 0.01''$.}
\label{fig:f2} 
\end{figure}

The v=1 masers are unresolved in 0.5\arcsec\ resolution images. 
Nevertheless, the high signal to noise ratio of the maps, in excess of
5000:1 on the strongest maser features, allows us to fit the {\it
centroid} of the maser position in each velocity channel to an accuracy of $\pm 0.01$\arcsec. 
The centroids are arranged in an 0.15\arcsec\ diameter ring with a
stable, systematic velocity pattern \citep{Plambeck90, Wright95,
Baudry98}.  In Figure~\ref{fig:f2}, the polarization vectors for
individual 1 \kms\ velocity channels are plotted on the corresponding
centroid positions for 3 different epochs.  Variations in the
polarization angle do not appear to correlate with movements of spots. 
Below we argue that the polarization angles of the v=1 masers may be
determined by the maser beaming direction, rather than the magnetic
field, accounting for the time variability. 

\citet{Plambeck90} were able to fit the v=1 maser spectrum and centroid
positions with a model of maser emission from a rotating, expanding
disk.  Higher resolution, VLBA observations \citep{Greenhill98,
Doeleman99} show, however, that the strongest maser features are
clustered into four groups, plausibly in the walls of the bipolar
outflow from source~I.  Although Doeleman et al.  emphasize the
differences between the disk and outflow models, the distinctions are in
part semantic -- the disk modeled by Plambeck et al.  had a sharp inner
radius of 40 AU and a thickness of 40 AU, and the maser pump rate and
SiO density were fitted to obtain particularly strong maser emission
from the inner walls of this cylinder, not so different from the outflow
model.  Nevertheless, it is clear that the velocity field hypothesized
by Plambeck et al.  is an oversimplification.

\subsection{The v=0 Masers}

\begin{figure}
\begin{center}
\epsfig{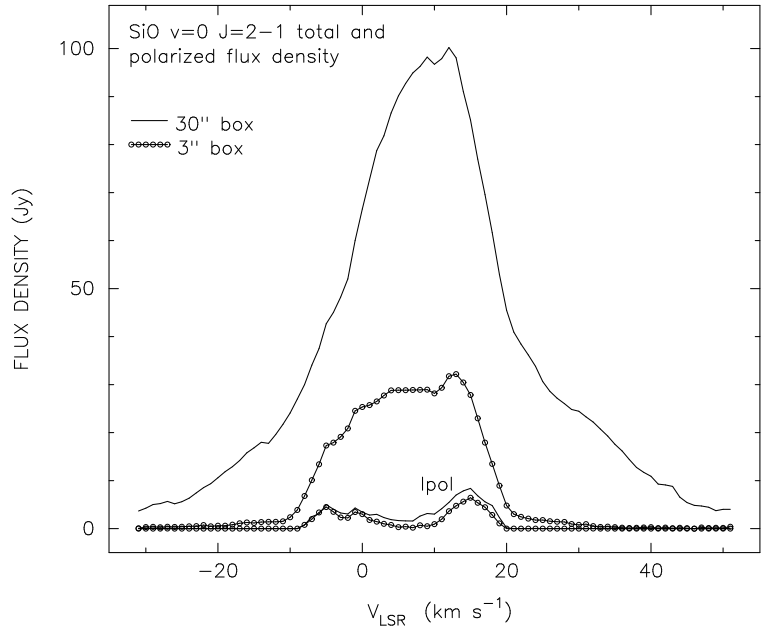}
\end{center}
\caption{Total intensity and polarized intensity v=0 \jtwo\ SiO spectra
integrated over $30'' \times 30''$ and $3'' \times 3''$ boxes centered
on Orion-IRc2.  Spectra for the $30''$ box were generated from BIMA C
array data taken in 2001 March, with a $10.4'' \times 6.7''$ synthesized
beam; for the $3''$ box, from A and B array data taken in 2000 December
and 2001 February, with a $0.97'' \times 0.50''$ beam.  The polarized
intensities (two lower curves) are similar, indicating that most of the
polarized flux originates in the $3''$ inner region.  }
\label{fig:f3}
\end{figure}

The bulk of the v=0 \jtwo\ SiO emission from the Orion-KL core is
thermal.  Much of it originates from high velocity gas in the bipolar
outflow from IRc2 \citep{Wright83}.  With arcsecond angular resolution,
most of this extended emission is resolved out, revealing a bright
``bowtie'' centered on source~I.  The peak brightness temperatures in
the bowtie are $> 2000$~K, almost certainly indicating maser emission
\citep{Wright95}.  Figure~\ref{fig:f3} compares total intensity and
polarized spectra of Orion-IRc2 generated from high and low resolution
maps.  The {\it polarized} flux densities integrated over $3'' \times
3''$ and $30'' \times 30''$ boxes are nearly identical, indicating that
most of the polarized emission originates from the masers. 

\begin{figure}
\centering
\includegraphics[width=14cm]{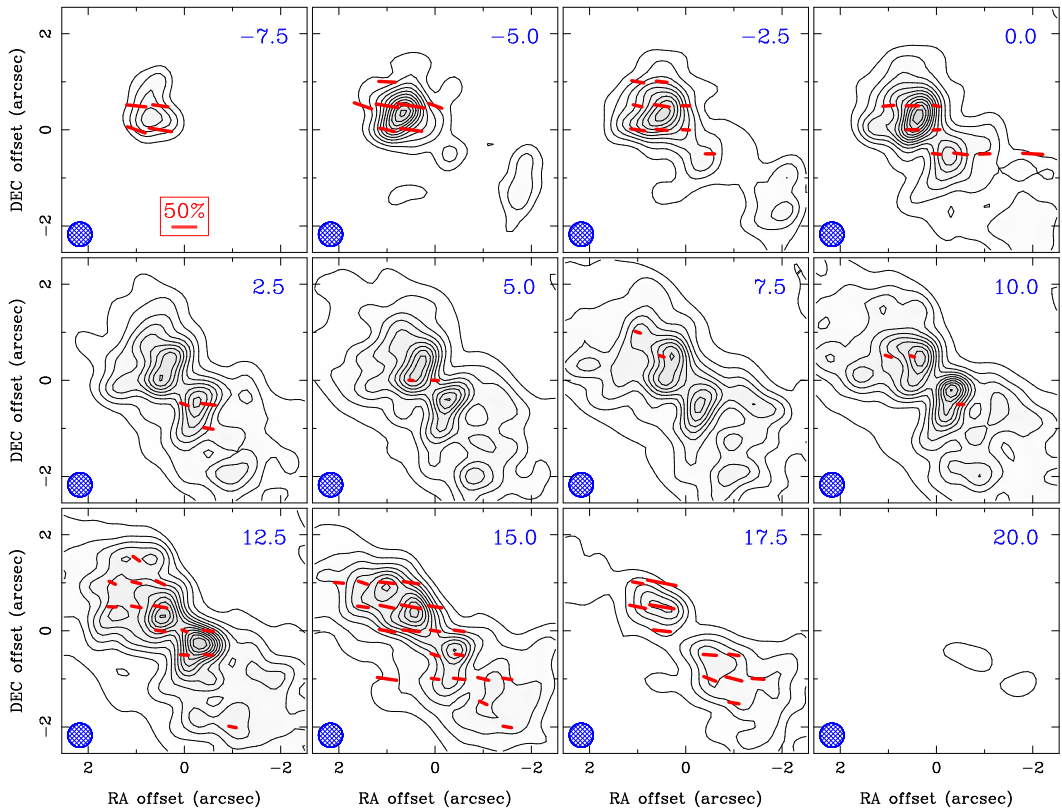}

\figcaption{Channel maps of \jtwo\ v=0 SiO emission in Orion-IRc2,
generated from BIMA A and B array data obtained in 2000 December and
2001 February.  The total intensity is indicated by contours, running
from 0.5~Jy/beam (325~K) to 3.25~Jy/beam (2100~K) in steps of
0.25~Jy/beam.  Polarization vectors are overlaid wherever the polarized
flux density $P > 0.2$~Jy/beam and, simultaneously, the total intensity
$I > 0.5$~Jy/beam.  The maximum fractional polarization is $\sim$50\% in
the $-5$ and $+17.5$ \kms\ maps; it is much weaker at line center. 
Offsets are relative to radio source~I at $\alpha = 5^h35^m14$\fs$505$,
$\delta = -5\arcdeg 22\arcmin 30$\farcs45 (J2000).  The LSR velocity, in
\kms, is shown at the upper right of each panel, the $0.5''$ FWHM
synthesized beam at the lower left. 
}

\label{fig:f4}
\end{figure}

SiO channel maps with $0.5''$ resolution are shown in
Figure~\ref{fig:f4}.  The r.m.s.  noise in the $Q$ and $U$ Stokes
intensity maps is $\sim$0.05~Jy/beam.  Polarization vectors are overlaid
wherever the linearly polarized flux $P = (U^2 + Q^2)^{1/2}$ is detected
with $4\sigma$ or greater significance; hence the polarization position
angles shown are uncertain by $\Delta\phi = \Delta P/(2P) < 7$\degr. 
Apart from a slight twist of the polarization vectors at the northern
edge of the \vlsr~12.5~\kms\ map, there are no obvious systematic
changes in polarization angle with velocity or position.  The
polarization direction, P.A.  $\sim$ 80\degr, is offset by 30\degr\ from
the long axis of the bowtie.  The fractional polarization is as high as
50\% in the channels near the upper and lower edges of the line profile,
near \vlsr~$-5$ and $+15$ \kms. 

The SiO linewidths in the bowtie are too large for this gas to be
gravitationally bound to source~I.  The ``shell'' H$_2$O masers in
Orion-IRc2 overlie the inner regions of the bowtie.  Proper motion
measurements show that these masers are moving outward from source~I at
$\sim$20~\kms, perpendicular to the axis of the high velocity outflow
\citep{Greenhill98}.  Probably the v=0 SiO masers and H$_2$O masers are
part of a low velocity outflow which is pushing into, or around, dense
gas in the equatorial plane of the star, plausibly the remnants of the
``pseudodisk'' which formed in the intial collapse of the cloud
\citep{Galli93}. 

\section{Discussion}

\subsection{Maser Polarization Mechanisms}

It is remarkable that a magnetic field with a strength of a few
milligauss can affect an SiO maser in any observable way, given that the
Zeeman splitting $g\Omega$ induced by the field is orders of magnitude
smaller than the maser linewidth $\Delta\nu$.  For SiO, $g\Omega/2\pi
\sim 0.2$ Hz/milligauss \citep{Nedoluha90}, while individual \jtwo\
maser features have linewidths of order 0.5 \kms, or $1.4 \times 10^5$
Hz.  Nevertheless, polarization arises because of the selection rules
which govern dipole transitions between energy levels.  Radiation
polarized parallel to the field cannot change the component of molecular
angular momentum $m$ along the field direction, hence $\Delta m = 0$;
conversely, $\Delta m = \pm 1$ for radiation polarized perpendicular to
the field \citep{Townes75}.  These rules cause the maser gain to differ for polarizations
parallel and perpendicular to the magnetic field. 

A simple example given by \citet{Goldreich73} illustrates this
mechanism.  Consider a maser in the \jone\ transition.  In the presence
of a magnetic field, the J=1 state is split into 3 sublevels
($m=+1,0,-1$).  If a maser beam propagates at right angles to the field,
radiation polarized parallel to the field induces transitions only from
the $(J,m) = (1,0)$ state to the $(0,0)$ state, while radiation
polarized perpendicular to the field induces transitions from both the
$(1,+1)$ and $(1,-1)$ states to the $(0,0)$ state.
If the pumping rates into the 3 upper sublevels are
equal, and if the maser is saturated -- that is, every excitation to the
upper state leads to the emission of a maser photon -- then two photons
polarized perpendicular to the field are emitted for every one photon
polarized parallel to the field.  Hence the net fractional linear
polarization is 1/3, perpendicular to the field. 

In the more general situation where the maser propagates at angle $0 <
\theta < \pi/2$ with respect to the field, radiation polarized parallel
to the projected field direction couples to all 3 magnetic sublevels,
while that polarized perpendicular to the field couples only to the
$m=\pm 1$ sublevels.  Hence, as $\theta$ decreases from $\pi/2$ toward
0, the gain of the parallel component increases while the gain of the
perpendicular component decreases.  In the limit $g\Omega \gg R \gg
\Gamma$, where $R$ is the stimulated emission rate and $\Gamma$ is the
decay rate from collisions or spontaneous emission, \citet{Goldreich73}
show that the net linear polarization for a \jone\ maser is
perpendicular to the projected field for $\theta > 54.7$\degr\
($sin^2\theta > 2/3$), and parallel to the projected field for $\theta <
54.7$\degr. 

\citet{Western84} found that these same asymptotic limits applied to the
\jtwo\ rotational transition.  In addition, they considered the case
where the magnetic sublevels were pumped unequally, which can easily
occur if the vibrational energy levels are excited by an anisotropic
radiation field, for example by infrared photons from a nearby star. 
Small differences in the absolute populations of these sublevels
correspond to large fractional differences in the population inversion,
and hence the maser gain.  Therefore, anisotropic pumping tends to
enhance the fractional polarization, and also changes the angle $\theta$
at which the polarization flips from perpendicular to parallel. 
Nonetheless, as long as $g\Omega \gg R$ and $\Gamma$, the net linear
polarization is either parallel or perpendicular to the projected field
direction. 

Unhappily, the requirement $g\Omega \gg R$ is likely to be violated for
intense masers, for which the stimulated emission rate $R$ becomes
large.  As $R$ increases, the ``good'' quantum axis shifts from the
magnetic field direction to the propagation direction of the maser beam. 
Calculations by \citet{Nedoluha90} show that the position angle of the
linear polarization is neither parallel nor perpendicular to the
projected magnetic field in this case, over a broad range of maser
intensities.  For example, Figure 3 of Nedoluha \& Watson shows that the
polarization position angle for a \jtwo\ SiO maser propagating at
15\degr\ to the magnetic field direction shifts from 20\degr\ for
$R=g\Omega$, to 80\degr\ for $R=100\,g\Omega$.  As the maser intensity
increases and $R$ exceeds $g\Omega$, the calculations also show that the
fractional polarization decreases. 

\subsection{The v=1 Masers}

The v=1 SiO masers near IRc2 are sufficiently intense that the
stimulated emission rate $R$ almost certainly is larger than the Zeeman
splitting $g\Omega$.  The brighter \jone\ maser spots in the VLBA map of
\citet{Doeleman99} have flux densities of 1~Jy in an 0.2 milliarcsecond
beam, corresponding to brightness temperatures of $2 \times 10^{10}$ K. 
To estimate the stimulated emission rate $R$ for an SiO molecule near
the surface of such a maser, one must guess the solid angle $d\Omega$
into which the maser radiation is beamed.  Taking $d\Omega = 10^{-2}$
steradians,
\begin{equation}
R=BU={{8\pi^3\mu^2}\over{3h^2}} {{(J+1)}\over{(2J+3)}}\ 
     {I}\, {{d\Omega}\over{c}} \sim 30\ s^{-1}
\end{equation}
where $B$ is the Einstein B coefficient for the $J+1 \rightarrow J$
transition, $U$ is the radiation energy density in
ergs~cm$^{-3}$~Hz$^{-1}$, $\mu= 3.1 \times 10^{-18}$ esu-cm is the
dipole moment for SiO, and $I = 2kT/\lambda^2$ is the maser's specific
intensity.  For comparison, the spontaneous radiative decay rate from
the v=1 to the v=0 vibrational level is 5 s$^{-1}$ \citep{Hedelund72}
and the \mbox{SiO--H$_{2}$} collision rate in 1000~K gas with
density $10^9$~\cucm\ is $\sim$1~s$^{-1}$, using the total rate
coefficient $k = 1.4 \times 10^{-9} (T/T_0)^{0.39}$~cm$^3$ s$^{-1}$
($T_0 = 2000$~K) given by \citet{Bieniek83}.  Since $R \gg \Gamma$, the
maser is saturated. 

In order to fulfill the requirement $g\Omega \gg R$, magnetic fields
$\ga$300~mG would be required.  For a 300~mG field the magnetic pressure
$B^2/8\pi$ is 25 times greater than the thermal pressure $nkT$, which
seems unlikely.  Thus, we reach the unwelcome conclusion that the
polarization direction of the v=1 masers is a poor indicator of the
magnetic field direction near IRc2.  The model calculations of
\citet{Nedoluha90} show that the polarization position angle is a
function of $R/g\Omega$, and hence of the maser flux density.  Thus, one
can understand both (1) the variations in polarization angle which occur
across the maser line profile, and (2) the changes in polarization
direction which occur on time scales of years as the maser intensities
change.  For $R/g\Omega > 1$ the models also suggest that the most
intense masers will have the lowest fractional polarization, as observed.

An observational result of note is the maser feature with 30\%
fractional polarization which appeared at \vlsr $\sim 7$ \kms\ in 1999
June.  Such high fractional polarization is difficult to achieve unless
the masers are anisotropically pumped by infrared radiation from the
central star \citep{Western84}.  Anisotropic pumping also can change the
critical angle at which the polarization becomes parallel or
perpendicular to the field, perhaps explaining the anomalous position
angle observed for this feature.  Anisotropic pumping can produce
strongly polarized masers even in the absence of a magnetic field
\citep{Western83}; this mechanism is thought to explain the tangential
polarization pattern seen the VLBA maps of SiO masers around evolved
stars \citep{Desmurs00}. 

\subsection{The v=0 Masers}

In contrast to the situation for the v=1 masers, the SiO masers in the
ground vibrational state almost certainly {\it do} fulfill the
requirement that $g\Omega \gg R$ and $\Gamma$, hence their polarization angles
should be perpendicular or parallel to the magnetic field. 

The v=0 maser emission region is extended with respect to our
synthesised beam, with peak brightness temperatures of about $2000$~K. 
To compute an upper limit for the stimulated emission rate, we assume
that the masers radiate isotropically, so the radiation energy density
$U = (4\pi/c)(2kT/\lambda^2)$.  Then $R = BU = 1.4 \times 10^{-2}$
s$^{-1}$.  The spontaneous radiative rate $A_{21} = 2.9 \times 10^{-5}$
s$^{-1}$ is relatively slow, so the decay rate $\Gamma$ from the J=2
level is set by the SiO-H$_{2}$ collision rate.  Modeling by
\citet{Zeng87} predicts that v=0 SiO masers originate in gas with
density $10^7$ \cucm.  If the gas kinetic temperature is $\sim$300~K,
the rate coefficient given by \citet{Bieniek83} implies a collision rate
of $6 \times 10^{-3}$ s$^{-1}$.  For a 1~mG field the Zeeman frequency
$g\Omega/2\pi \sim 0.2$ s$^{-1}$.  Since $g\Omega \gg R > \Gamma$, the
v=0 masers should be excellent tracers of the magnetic field direction. 
Reassuringly, the position angle of the SiO v=0 polarization appears to
be stable with time.  The maps in Figure~\ref{fig:f4} were made from
data taken in 2000 December and 2001 February; the position angle is the
same, P.A.  $\sim$80\degr, in lower quality images from 1996, 1997, and
1999.  We note that this is close to the median position angle of the
v=1 masers. 

\begin{figure}
\centering
\includegraphics[width=10cm]{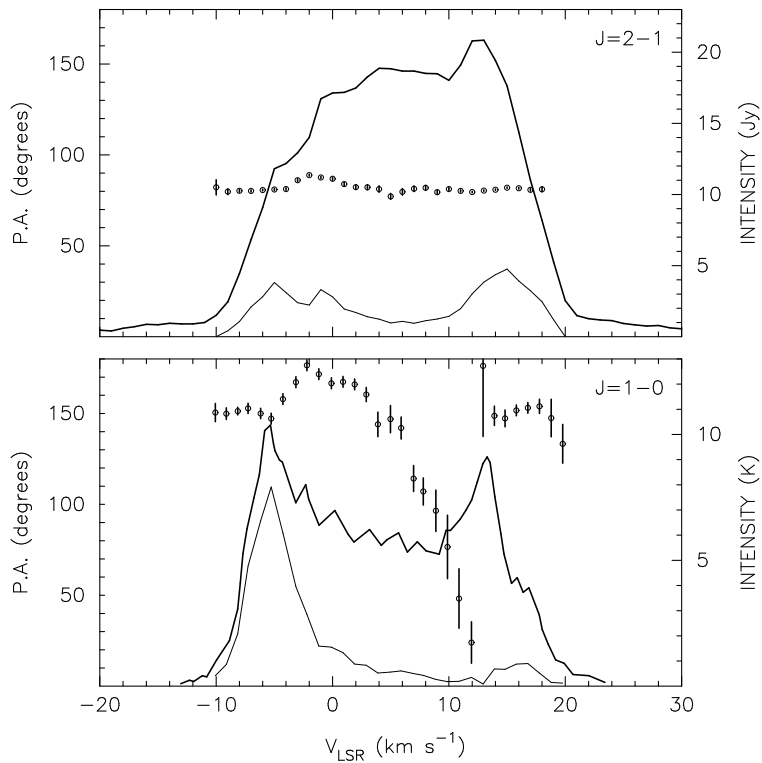}

\figcaption{Total intensities (thick curves), polarized intensities (thin
curves) and polarization position angles for the \jtwo\ and \jone\ SiO
masers in the v=0 vibrational state, toward Orion-IRc2.  The \jtwo\
transition was observed interferometrically with BIMA (this paper).  To
average over the maser emission region, we convolved channel maps of the
I, Q, and U Stokes parameters with a 3\arcsec\ FWHM Gaussian, then
computed the fluxes and position angles toward source~I from these
smoothed maps.  The \jone\ data \protect\citep{Tsuboi96} were obtained with the
Nobeyama 45-m telescope with a $39''$ beam.  The position angles
measured for the \jtwo\ and \jone\ transitions differ by roughly
70\degr.}

\label{fig:f5}
\end{figure}

One glaring discrepancy complicates the analysis: from observations made
in 1994, \citet{Tsuboi96} measured polarization position angles
P.A.~$\sim$150\degr\ for the \jone\ masers in the v=0, 1, and 2
vibrational levels.  As shown in Figure~\ref{fig:f5}, these position
angles differ by roughly 70\degr\ from the values we measure for the
\jtwo\ line.  Given that the position angle we measured for the v=0
\jtwo\ line was stable for 4 years, we doubt that this discrepancy can
be attributed to time variability.  It's also unlikely that the
39\arcsec\ beam used for the \jone\ measurements picked up a halo of
emission polarized almost perpendicular to the core, since we find that
virtually all of the polarized \jtwo\ flux originates in the central
$3''$ region.  This leaves two possibilities: either (1) the \jtwo\ and
\jone\ masers have intrinsically different polarizations; or (2) the
intrinsic polarizations are identical, but the observed position angles
are twisted by Faraday rotation.  Possibility (1) is hard to rule out. 
However, in the grid of theoretical calculations done by
\citet{Nedoluha90} there appear to be no cases where the the \jtwo\ and
\jone\ polarization position angles differ by more than 20\degr.  And,
if the \jtwo\ position angle we measure is parallel or perpendicular to
the magnetic field, then the field is aligned neither with the bipolar
outflow from IRc2 (P.A.$~\sim$145\degr, as defined by the VLBA
observations of the SiO masers), nor with the axis of the v=0 disk
(P.A.~$\sim$50\degr). 

\subsection{Faraday rotation}

We now explore the possibility that Faraday rotation causes the
discrepancy between the \jtwo\ and \jone\ polarization position angles. 
Faraday rotation occurs when linearly polarized emission propagates
through an ionized plasma with a magnetic field component along the line
of sight.  The plane of polarization is rotated by $\theta =
\lambda^2RM$, where the rotation measure is given by $RM = 8.1 \times
10^5 \int n_e B_{\parallel} \, dl$ rad~m$^{-2}$.  Here $n_e$ is the
electron density in \cucm, $B_{\parallel}$ is the component of the
magnetic field along the line of sight in gauss; and $dl$ is in parsecs. 
If the propagation vector is along the magnetic field direction, then
the plane of polarization rotates in a clockwise direction, following
the right hand rule.  The minimum rotation measure which can bring the
\jone\ and \jtwo\ position angles into agreement is $RM = 3.3 \times
10^4$~rad~m$^{-2}$, which rotates the \jtwo\ ($\lambda$3.5 mm) position
angle by 23\degr, and the \jone\ ($\lambda$7 mm) position angle by
92\degr. 

Because the Faraday rotation must be reasonably uniform across the
$\sim$3\arcsec ~maser emission region, we presume that it occurs in the
foreground Orion~A \ion{H}{2}\ region (M42), not in ionized gas local to
IRc2 -- e.g., on the surface of a circumstellar disk.  What is the
plausible rotation measure through Orion~A? We estimate the electron
column density from the work of \citet{Wilson87}, who modeled the
\ion{H}{2}\ region as a series of 9 concentric cylindrical slabs in
order to fit radio continuum and recombination line data.  The line of
sight to IRc2 passes through slabs 3--9, for which the model column
density $\int n_e dl \sim 650$ \cucm\,pc.  We take the line of sight
component of the magnetic field from the work of \citet{Troland89}, who
measured the 21 cm Zeeman effect in the sheet of neutral absorbing gas
in front of Orion~A.  The inferred field is reasonably uniform, with no
reversals of direction; the value toward IRc2 is approximately
$-50\mu$G.  The negative sign indicates that the field lines point {\it
toward} us.  Then the rotation measure is {\it positive}, meaning that
the observed position angle {\it increases} with wavelength
\citep{Heiles87}, which is what we are hoping to find.  If the magnetic
field is the same inside the \ion{H}{2}\ region, then $RM = 2.6 \times
10^4$ rad~m$^{-2}$, within 20\% of the value which brings the position
angles into concordance. 

The estimated rotation measure is comparable with the the upper limit
$RM < 2.8 \times 10^4$~rad~m$^{-2}$ deduced by \citet{Rao98} from the
agreement of the polarization position angles for dust emission at
wavelengths of 1.3~mm and 3.5~mm.  We note that the majority of these
dust polarization detections were made west and north of IRc2, farther
from the center of the \ion{H}{2}\ region. 

\subsection{Magnetic Field Direction}

\begin{figure}
\centering
\includegraphics[width=15cm]{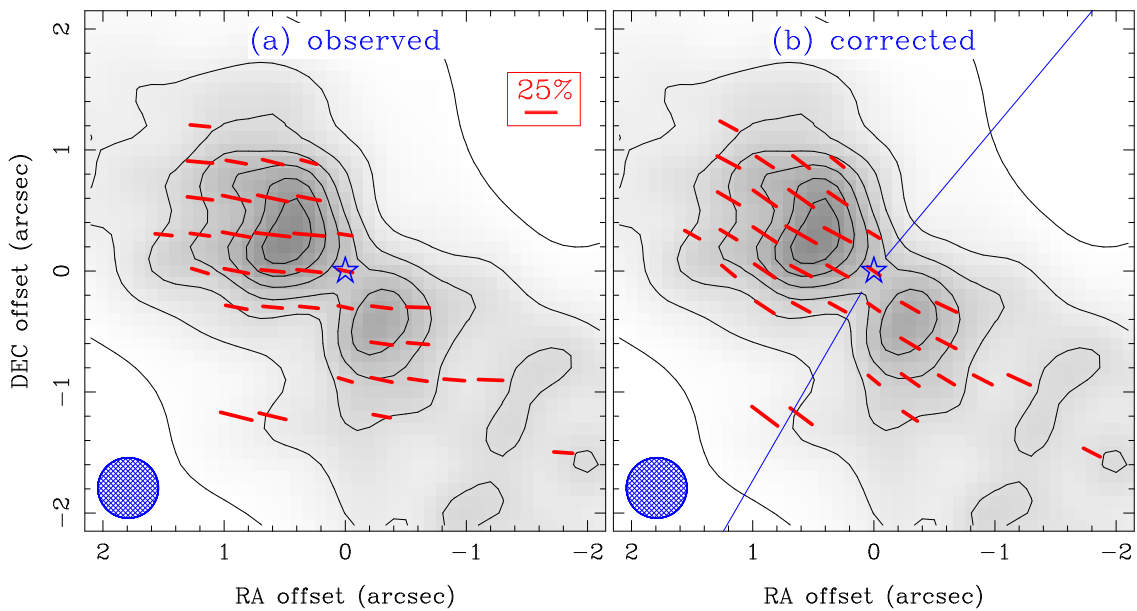}

\figcaption{(a) Polarization map of v=0 \jtwo\ SiO emission in Orion-KL,
integrated over a 28~\kms\ wide velocity range centered at \vlsr~5 \kms. 
The contour interval is 0.3~Jy/beam, or 200~K.  Vectors indicate the
polarization direction wherever there is a $\geq$4~$\sigma$ detection of
the polarized flux, corresponding to a position angle uncertainty of
$\pm 7$\degr.  The highest fractional polarization is 28\%.  The
position of radio source~I is denoted by an open star at the center of
the map.  (b) Same, except the polarization vectors have been
rotated by $-23$\degr\ to correct for Faraday rotation, as discussed in
the text.  Thin lines at P.A.  $-40$\degr\ and 150\degr\ indicate the
axes of the high velocity outflow from source~I, as defined by the cones
of v=1 masers close to source~I \protect\citep{Greenhill98}.  }

\label{fig:f6}
\end{figure}

Suppose we accept the hypothesis that Faraday rotation has twisted the
\jtwo\ position angle by $+23$\degr.  As shown in Figure~\ref{fig:f6},
this means that the intrinsic polarization angle is 57\degr, roughly
parallel with the plane of the SiO ``disk,'' and perpendicular to the
axis of the high velocity outflow from source~I.  The inclination angle
of the disk is uncertain, but the 3:1 axial ratio of both the SiO and
H$_2$O maser distributions suggest that its axis is tilted by $\la
20$\degr\ from the plane of the sky, that is, that the disk is seen
nearly edge-on. 

The magnetic field can be parallel or perpendicular to the SiO
polarization vectors -- but which is more likely? If the maser radiation
travels more or less parallel to the magnetic field, then it should be
polarized parallel to the projected field direction.  Thus, our
measurements are consistent with a magnetic field in the plane of the
disk, as long as the field is oriented roughly along the line of sight. 
Obviously, such a field cannot be azimuthally symmetric (e.g.,
ring-shaped or spiral) around IRc2.  If it were, the field would become
perpendicular to the line of sight in the outer parts of the bowtie, and
the polarization direction would flip by 90\degr. 

Our data are more easily reconciled with a field that is everywhere
perpendicular to the plane of the disk.  In this case the maser
radiation propagates at an angle of ~70\degr\ to the field, so the
polarization should be perpendicular to the field.  Then the field, at
P.A.  145\degr, is parallel to the axis of the high velocity outflow and
to the large scale Orion field mapped by far infrared observations
\citep{Schleuning98}.  The infrared measurements of
\citet{Chrysostomou94} and \citet{Aitken97} give a similar magnetic
field direction toward source~I, along the SE edge of the IRc2 infrared
peak.  Our data provide no convincing evidence that the field has been
pinched or twisted within $\sim700$ AU of source~I.  The twist seen in
the infrared data occurs on slightly larger scales.

\section{Conclusions}

We attempted to infer the magnetic field morphology toward Orion-IRc2
from polarization measurements of the v=0 and v=1 \jtwo\ SiO masers
associated with this star.  The polarization vectors are expected to be
perpendicular or parallel to the projected magnetic field direction
provided that the Zeeman splitting $g\Omega \gg R$ and $\Gamma$, where $R$ and
$\Gamma$ are the stimulated emission and decay rates from the maser's
upper rotational level. 

We find that the polarization position angles of the v=1 masers vary on
time scales of years.  These intense masers, which originate only
$\sim$40 AU from the star, probably are not suitable indicators of the
magnetic field direction because the stimulated emission rate exceeds
the Zeeman splitting.  Typically their fractional polarization is a
few percent, although in 1999 a weak feature with 30\% polarization
appeared.  

The v=0 masers are much weaker and should be good probes of the magnetic
field direction.  These masers originate \mbox{100--700~AU} from the
star in an elongated, bowtie shaped region, possibly along the top and
bottom surfaces of an underlying flared disk.  Maps of the v=0 emission
with 0.5\arcsec\ (250~AU) angular resolution show that, contrary to our
expectations: (1) the polarization vectors within $\sim$700 AU of the
star are straight, with no evidence of a twist or hourglass morphology;
(2) the measured polarization position angle of 80\degr\ is neither
parallel nor perpendicular to the high velocity outflow from IRc2; and
(3) the \jtwo\ polarization position angle differs by 70\degr\ from the
\jone\ position angle previously measured by \citet{Tsuboi96}. 

The discrepancy between the \jtwo\ and \jone\ position angles may be
attributable to Faraday rotation by plasma in the foreground Orion~A
\ion{H}{2}\ region.  A rather high, but not implausible, rotation
measure $RM = 3.3 \times 10^4$~rad~m$^{-2}$ is required to bring the two
position angles into agreement.  Then the intrinsic polarization
direction is parallel to the plane of the disk and perpendicular to the
axis of the high velocity outflow from source~I.  Probably the
polarization vectors are perpendicular to the field in this case, so the
field threads the disk poloidally, parallel to the outflow and to the
large scale magnetic field in Orion.  The absence of a discernable pinch
or twist in the field near IRc2 may indicate that as the molecular cloud
collapsed the gas density in the pseudodisk exceeded $10^{11}$~\cucm, at
which point the magnetic field decouples from the infalling gas
\citep{Nakano86}.  Alternatively, the outflow or subsequent evolution of
the disk has erased all traces of the pinched field. 

The high fractional polarizations observed for the v=0 masers -- up to
50\% in our \jtwo\ data and up to 80\% in the \jone\ observations of
\citet{Tsuboi96} -- suggest that anisotropic excitation by infrared
photons play an essential role in pumping these masers.

\acknowledgements

This work was partially supported by NSF grant AST-9981308  to the
University of California.

\end{document}